\documentclass{sbrt2018eng}

\usepackage{amsmath}

\begin{document}

\title{Development and Evaluation of Least Square Satellite Tracking in Real Antenna Control System}

\author{Thiago Sarmento,
Andrey Nakamura,
Itelo Filho,
Lucas Correa, \\
Marcos Takeda,
Adalbery Castro,
Aldebaro Klautau
\thanks{Andrey Nakamura, Itelo Filho, Lucas Correa, Thiago Sarmento, Marcos
Takeda, Adalbery Castro and Aldebaro Klautau are with Federal University of
Para, Belem, Brazil (e-mails: \{thiago.sarmento, andrey.nakamura, lucas.correa,
marcos.takeda\}@itec.ufpa.br, \{itelo, adalbery, aldebaro\}@ufpa.br).}}
\maketitle

\markboth{XXXVI SIMPÓSIO BRASILEIRO DE TELECOMUNICAÇÕES E PROCESSAMENTO DE
SINAIS - SBrT2018, 16-19 DE SETEMBRO DE 2018, CAMPINA GRANDE, PB} {XXXVI
SIMPÓSIO BRASILEIRO DE TELECOMUNICAÇÕES E PROCESSAMENTO DE SINAIS - SBrT2018,
16-19 DE SETEMBRO DE 2018, CAMPINA GRANDE, PB}

\begin{abstract}

This work describes the implementation of a satellite tracking technique in a
real antenna control system. The technique uses least squares estimation to
search for the best antenna pointing position to receive the satellite beacon
signal tracked in the 2D region created by the motion axes. The work also
presents the technique implementation results in the real system to prove its
operation with the beacon signal intensity and the orbit obtained with the
search over time.

\end{abstract}


\section{Introduction}

The satellites are essential for global communications, enabling new
possibilities and applications never imagined before. However, like all
technologies, the satellite communications show challenges to be overcome. One
of these problems is in maintaining the antenna’s pointing in the best reception
position of the satellite signal. Satellite tracking techniques are widely used
in earth stations antenna control systems \cite{hawkins1988tracking} to improve
signal reception of near geosynchronous satellites. Most algorithms are based on
both search and optimization algorithms applied to antenna pointing angles and a
beacon signal. 

The Satellites have a beacon signal, serving as a reference for reception of
antennas. This signal is used by the algorithms for the tracking of its orbit,
and the satellites studied in this work are the geostationary orbits inclined
\cite{prussing1993orbital}. The objective of these tracking techniques is to
maximize the beacon signal reception with respect to the azimuth and elevation
angles. As satellites keeps moving, searches must be done several times a day.

%TODO revisão bibliográfica / estado da arte aqui

\section{Least Squares}
In~\cite{Laine2000} a least squares framework is devised by modeling the beacon
signal level near the optimal pointing angle as a 2D parabola. The algorithm
uses a square movement pattern to perform measurements. At each corner the
antenna stops by a predetermined time and then a beacon measurement is made and
stored, along with its azimuth and elevation. This data is then used to estimate
a 2D parabola by the least squares method.

The beacon model is given by the parabola equation
\begin{equation}
\label{eq:parabola}
L(x,y) = K_x ( x - p )^2 + K_y ( y - q )^2 + L_{pq}
\end{equation}
where $K_x$ and $K_y$ are the quadratic coefficients for azimuth and elevation,
respectively. $x$ and $y$ are the measurement angles, $p$ and $q$ are the
satellite direction, $L_{pq}$ is the maximum beacon signal and $L(x,y)$ is the
beacon level at pointing angle $(x,y)$.

By expanding \eqref{eq:parabola} one can write
\begin{multline}
\label{eq:expanded}
L(x,y) - K_x x^2 - K_y y^2 = \\
-2 K_x x p -2 K_y y q + L_{pq} + K_x p^2 + K_y q^2
\end{multline}
and then defining
\begin{align}
\beta^T &= \left[\begin{array}{ccc}
-2 K_x p & -2 K_y q & (L_{pq} + K_x p^2 + K_y q^2)
\end{array}\right] \\
X_i &= \left[\begin{array}{ccc}
x_i & y_i & 1
\end{array}\right] \\
Y_i &= L(x_i,y_i) - K_x x_i^2 - K_y y_i^2
\end{align}
where, at each time instant, sensors provide samples $x_i$, $y_i$ and $L(x_i,
y_i)$ from azimuth resolver, elevation resolver and beacon receiver,
respectively.

We can then rewrite \eqref{eq:expanded} simply as
\begin{equation}
Y = X \beta
\end{equation}
where $Y$ is a $N \times 1$ matrix and $X$ is a $N \times 3$ matrix, with $N$
being the number of samples. Each row of $Y$ is $Y_i$ and each row of $X$ is
$X_i$, $i$ varying from $1$ to $N$. The least squares solution is then given by
the well-known equation
\begin{equation}
\beta = (X^T X)^{-1} X^T Y
\end{equation}
and we can retrieve the satellite position and beacon level by
\begin{align}
p &= -\frac{\beta_1}{2 K_x} \\
q &= -\frac{\beta_2}{2 K_y} \\
L_{pq} &= \beta_3 - K_x p^2 - K_y q^2 \text{.}
\end{align}

One drawback of this linear approach is that the values if $K_x$ and $K_y$ must
be known before hand. In~\cite{Laine2000}, $K_x$ is calculated as $K_x = K_y
\cos^2 (h_0)$, where $h_0$ is the elevation of the satellite position, but $K_y$
is still needed. For our setup we managed to estimate $K_y \approx -11.4$, and
$K_x$ is in the range $[-1.2 -1.0]$ depending on the actual elevation position.
%TODO figura do movement pattern

\section{Recursive least squares (RLS)}
The RLS algorithm can perform least squares in a filter form, avoiding matrix
inversion and adding a exponential weighting function which decreases the
weights for older samples. The advantage is that using a filter form, we can
estimate the parabola by storing only the filter state. Also, the weighting
function can help to model the satellite movement occurring while estimation is
performed.

The RLS algorithm is given by
\begin{align}
Q &= \frac{P_{i-1}}{\lambda + X_i P_{i-1} X_i^T} \\
K &= Q X_i^T \\
\hat{Y}_i &= X_i \beta_{i-1} \\
e &= Y_i - \hat{y}_i \\
\beta_i &= \beta_{i-1} + K e \\
P_i &= \frac{1}{\lambda} (I - K X_i) P_{i-1}
\end{align}
where $P$ is a positive-definite matrix, $x$ is a vector with input data,
$\beta$ is a vector with the estimated model coefficients and $\lambda$ is the
forgetting factor.

The forgetting factor $\lambda$ is a constant between $0$ and $1$ which defines
the weight of past samples in the estimation. Defining $\tau =
\frac{1}{1-\lambda}$ as the memory horizon of the RLS algorithm, samples older
than $\tau$ have small weights in the algorithm. For example, $\lambda=0.98$
gives $\tau=50$ samples. For $\lambda=1$, the memory horizon is infinite,
yielding the same estimation of the regular least squares, as every sample has
the same weight.

\section{Tracking algorithm}
Tracking is performed in a timed basis in order to minimize antenna movement,
reducing mechanical wear and tear. The tracking algorithm in this work has four
steps:
\begin{enumerate}
\item Perform a displacement pattern on the neighborhood of the actual position
to acquire data
\item Estimate the optimal position with acquired data using
\item Move to the estimated position
\item Wait for the next tracking cycle
\end{enumerate}

\begin{figure}[htb]
  \centering
  \includegraphics[width=0.5\columnwidth]{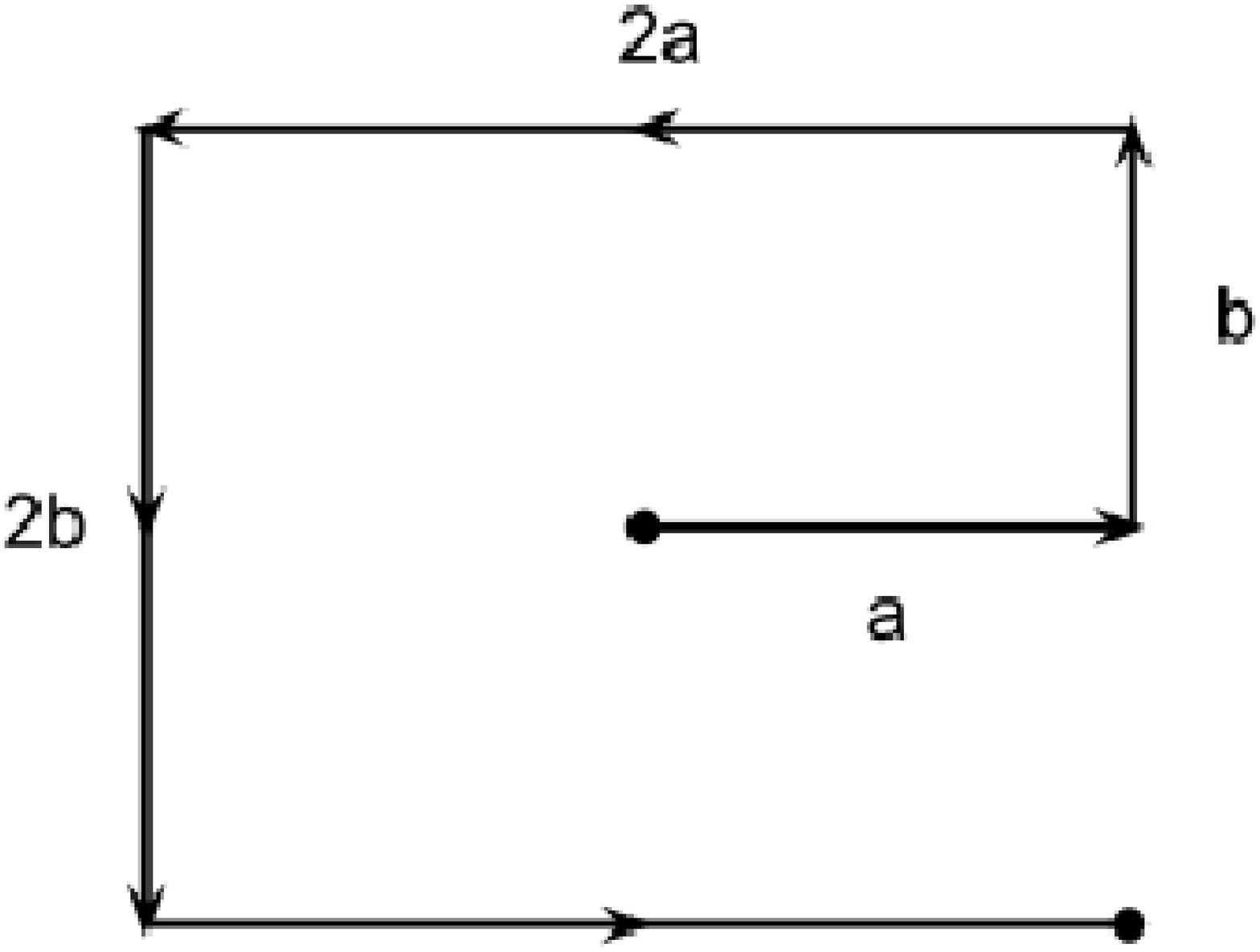}
  \caption{Displacement movement pattern. \label{fig:mov}}
\end{figure}

The displacement pattern used is mostly the same from~\cite{Laine2000} and
reproduced in Figure \ref{fig:mov}, the only difference is that in this work we
adapted to a rectangle instead of a square. This is because we observed
different variations on each axis, so we can adjust accordingly.

One key difference is that in~\cite{Laine2000} the measurements are only made at the
vertices of the square, but in out setup these measurements are taken every 20~ms, thus
increasing the number of samples needed to be stored by the conventional least squares,
while in RLS this is not a problem due to the filter form of the algorithm.

\section{Antenna Setup}

\begin{figure}[htb]
  \centering
  \includegraphics[width=0.8\columnwidth]{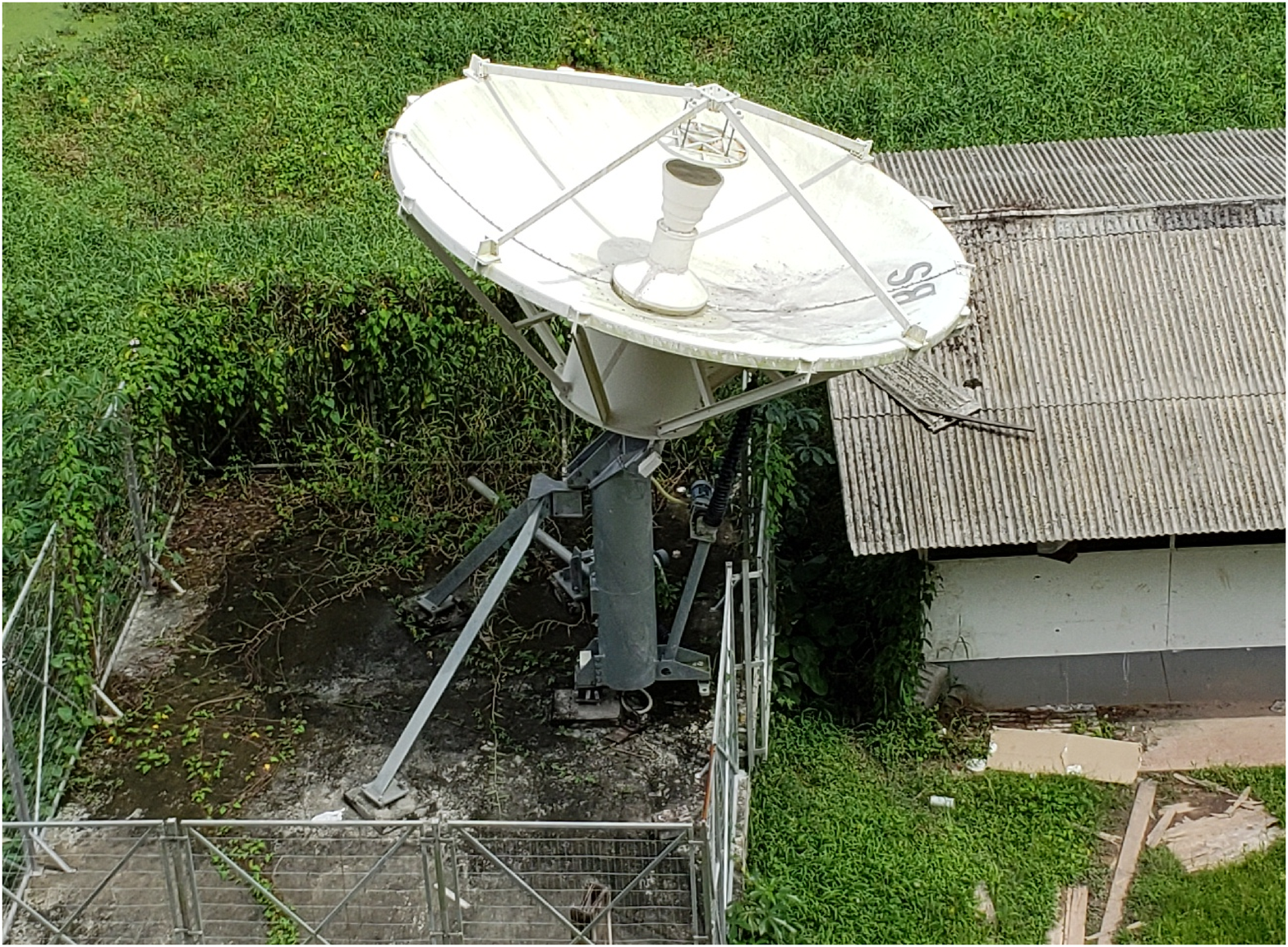}
  \caption{3 axes antenna installed in the Guamá STP. \label{fig:antenna_c}}
\end{figure}

%TODO describe LASSE's antenna setup, including a figure
The antenna used, shown in the Figure \ref{fig:antenna_c}, is parabolic
measuring 3 meters and having 3 axes of freedom (azimuth, elevation and
polarization), which is installed in the Guamá STP (Science and Technology Park)
belonging to LASSE (Telecommunications, Automation and Electronics Research and
Development Center).

The antenna receives the satellite signal, which passes the signal to LNB (Low
Noise Block) which makes RF to IF down conversion of signal, and thus transmits
this signal to the beacon receiver \cite{generalD} via cable. The received
signal from the beacon receiver is linearized to a voltage range of 0 to 10 V
which is directly proportional to the intensity of the satellite beacon signal.
This antenna is part of a project to develop a system to automatically
control the antennas movement to make possible the satellite tracking with the
techniques implementation.

\begin{figure}[htb]
  \centering
  \includegraphics[width=1\columnwidth]{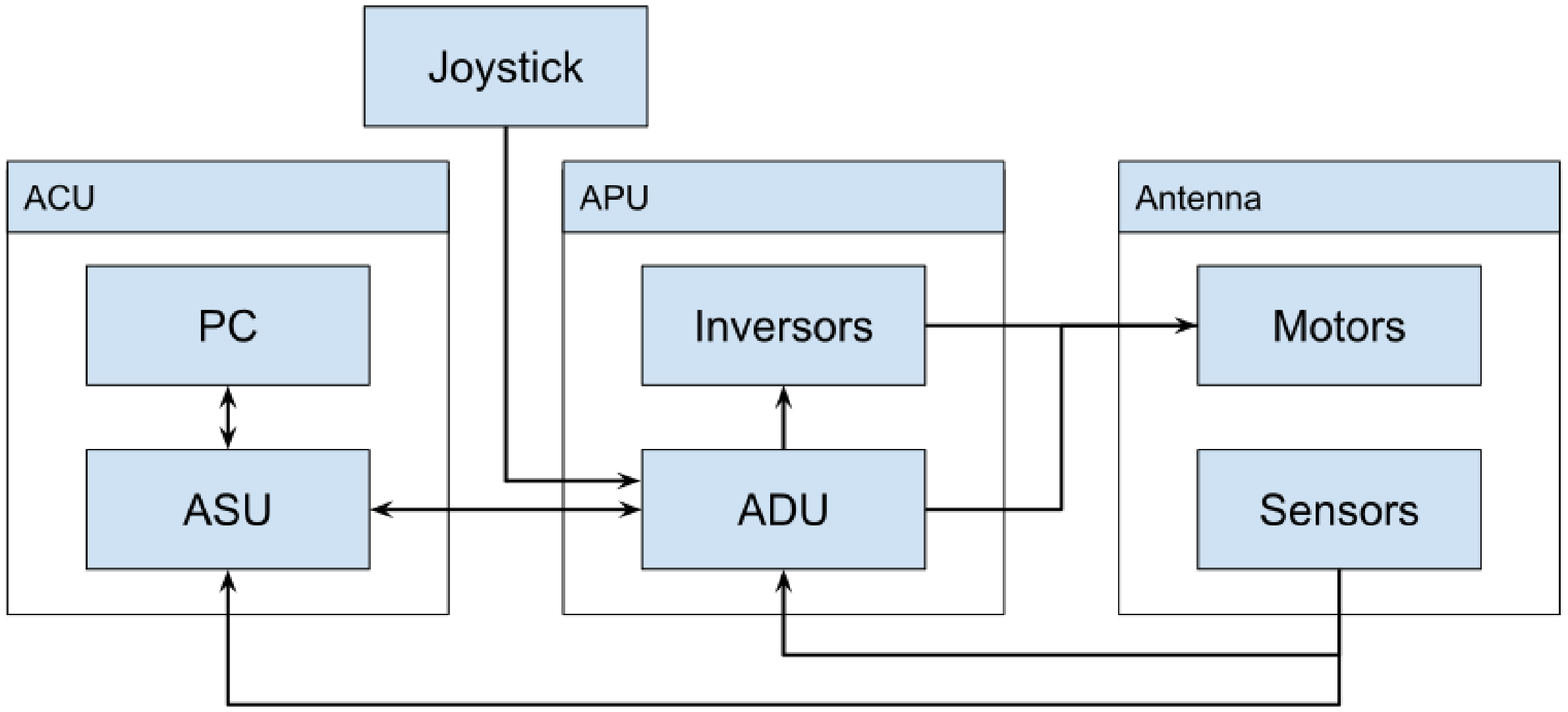}
  \caption{Antenna Monitoring and Control System overview. \label{fig:overview}}
\end{figure}

The antenna monitoring and control system is designed to control 3-axis antennas
(azimuth, elevation and polarization), controlling the antenna movement manually
or automatically. The system is also responsible for monitoring the beacon
signal strength, correcting the antenna pointing position for best reception,
using the tracking techniques developed to control the antenna. The system
consists of an antenna power unit (APU) and antenna control unit (ACU), the
layout of which can be seen in Figure \ref{fig:overview}.

The APU is responsible for activation and/or controlling the electrical
components of the system, controlling the movement of the antenna through the
azimuth, lift and polarization axis motors. The ACU has the system control
function, sending instructions to the APU and performing the available tracking
techniques, as well as storing settings and other important information such as
position register and beacon signal strength.

\subsection{Antenna Power Unit}
The antenna power unit is responsible for the control of the electrical
components of the system such as inverters and motors of the antenna, reading of
sensors installed in the antenna and the electrical board, communication with
the ACU and Joystick for receiving commands and sending of system status and
actions of that are taken in emergency situations.

\begin{figure}[htb]
  \centering
  \includegraphics[width=1\columnwidth]{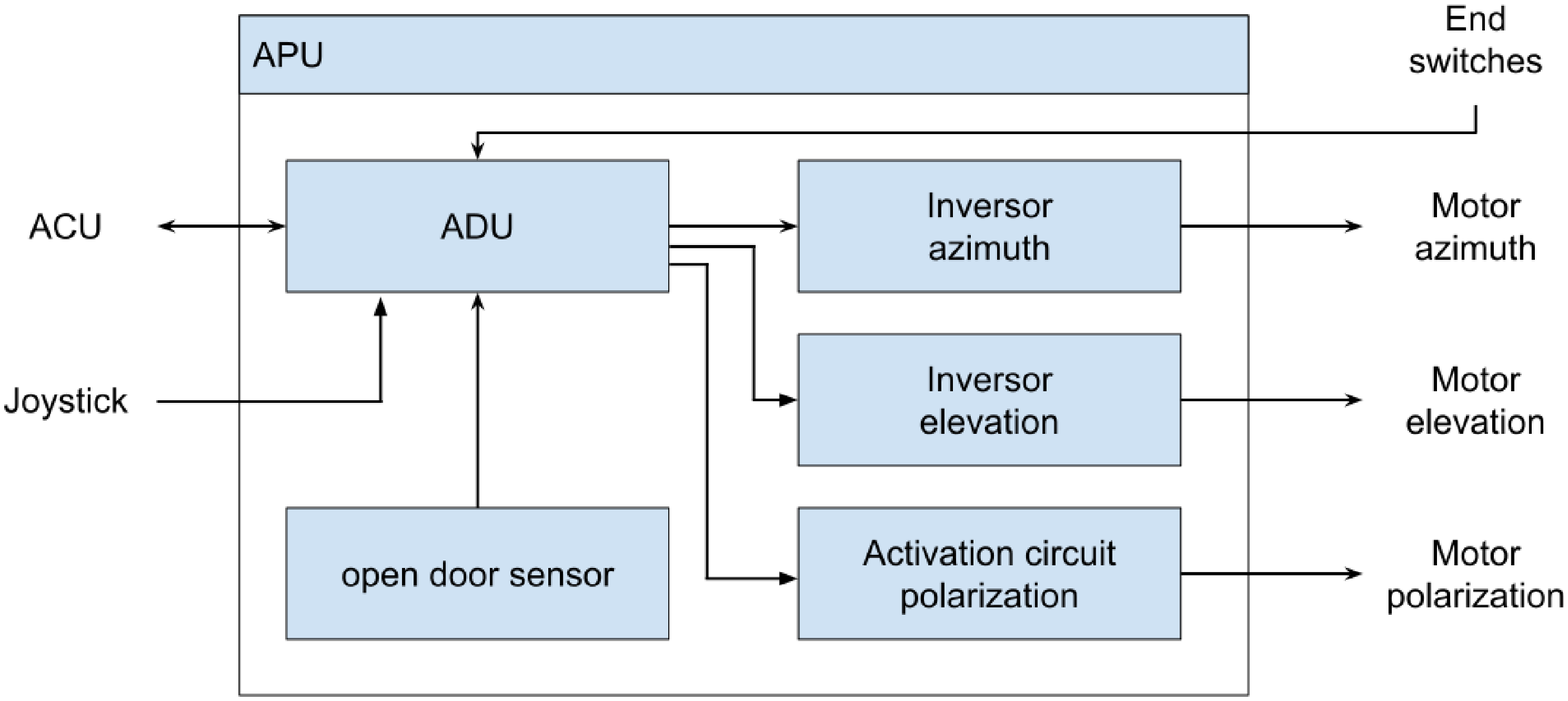}
  \caption{Antenna Power Unit diagram. \label{fig:APU}}
\end{figure}

The Antenna Drive Unit (ADU) was the prototype developed in project to control
the handling functions. The microcontroller \cite{at90can} used in ADU reads the
sensors connected to it, sends status to ACU and receives move instructions for
activating the antenna inverters and motors through the CAN (Controller Area
Network) interface.

\subsection{Antenna Control Unit}
The antenna control unit is responsible for the intelligence of the system,
sending instructions for moving to the APU, which can come from user interaction
through the graphical interface, already configured satellite tracking
techniques or emergency situations.

\begin{figure}[htb]
  \centering
  \includegraphics[width=1\columnwidth]{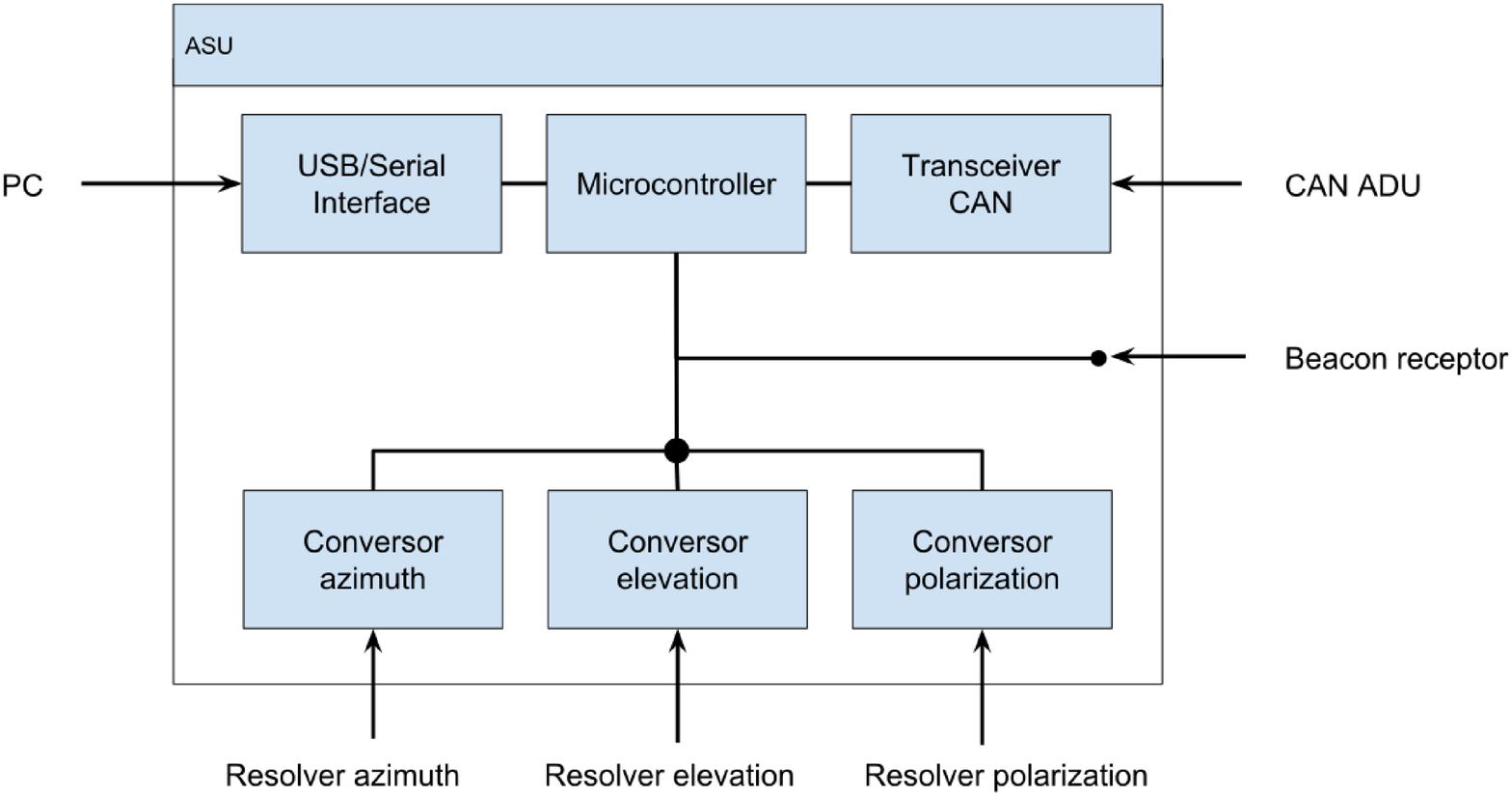}
  \caption{Antenna Control Unit diagram. \label{fig:ASU}}
\end{figure}

The computer is responsible for running software that monitors all system
operations, stores system settings, runs available trace techniques, and
receives user commands through the GUI (Graphical User Interface).
The ASU Antenna Sensors Unit was the prototype developed in the project to read
the sensors and promote ACU communication with the ADU. The data received by ADU
and the values read in the resolvers \cite{resolver,ad2s} and beacon receiver
sensors are sent to the PC and it responds with commands to be passed. The ASU
makes the connection between the computer located on ACU and APU, receiving the
messages and passing the commands. In order to communicate with ADU, ASU uses
the available CAN interface on its microcontroller, and to communicate with the
computer it has a USB/Serial converter to interface with. The microcontroller
must also read the sensors connected to ACU to pass the data to the computer.
All activities performed by ASU must be performed in real time, updating both
the PC with the data collected on system status and ADU on commands related to
the antenna movement.

\section{Tracking Results}

In our tests, our system tracked the Brasilsat B3 satellite \cite{BrasilsatB3}, which is in
inclined orbit. Running our system for approximately 24 hours, we obtained a
data log from the antenna control system containing information about azimuth,
elevation and beacon levels.

\begin{figure}[htb]
  \centering
  \includegraphics[width=1\columnwidth]{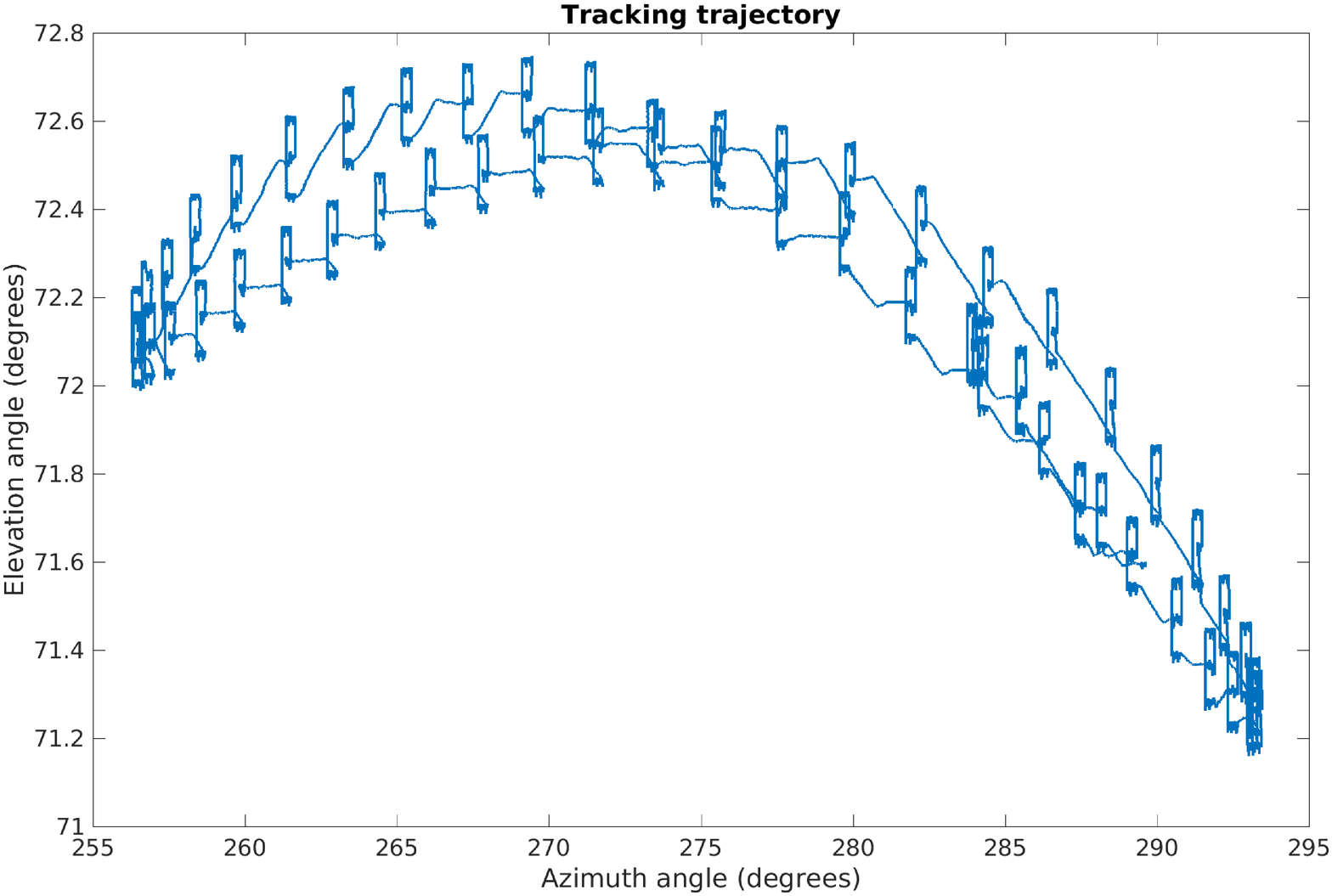}
  \caption{Tracking trajectory. \label{fig:aziele}}
\end{figure}

Figure \ref{fig:aziele} shows the tracking trajectory for our algorithm, showing
the movement pattern of the measurement phase, and then moving to the
maximizing position. The overall trajectory shape is, as expected, a figure 8
shape, frequently observed for inclined orbit satellites.

\begin{figure}[htb]
  \centering
  \includegraphics[width=1\columnwidth]{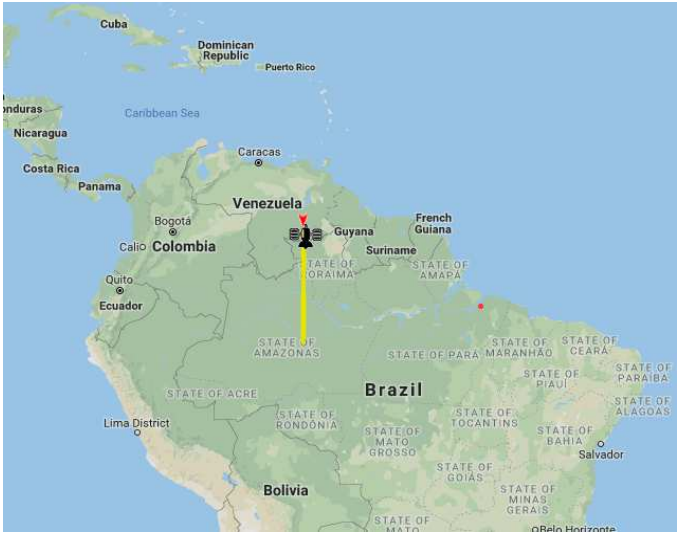}
  \caption{n2y0 orbit tracking \cite{website:n2yo}. \label{fig:n2yo}}
\end{figure}

\begin{figure}[htb]
  \centering
  \includegraphics[width=1\columnwidth]{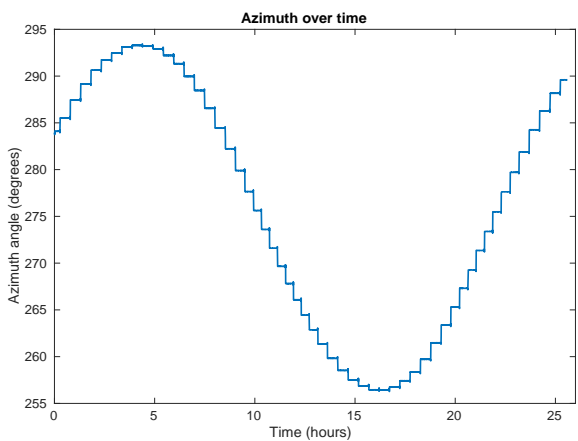}
  \caption{Azimuth angle over time. \label{fig:azi}}
\end{figure}

\begin{figure}[htb]
  \centering
  \includegraphics[width=1\columnwidth]{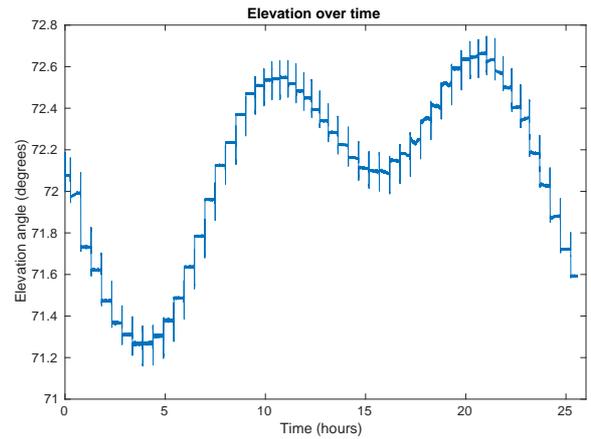}
  \caption{Elevation angle over time. \label{fig:ele}}
\end{figure}

Figure \ref{fig:azi} and \ref{fig:ele} shows the azimuth and elevation angles
over time. The discontinuities are caused by the tracking algorithm, which
adjust the antenna every tracking cycle.
We can note that the azimuth angle variation is greater than 30 degrees while
the variation in elevation angle is smaller than 2 degrees, this is because our
earth station location relative to the satellite. The inclined orbit makes the
satellite move in the north-south direction as seen in Figure \ref{fig:n2yo},
and as our earth station is near the equator line, most of the satellite
movement is in the azimuth angle of our antenna.

\begin{figure}[htb]
  \centering
  \includegraphics[width=1\columnwidth]{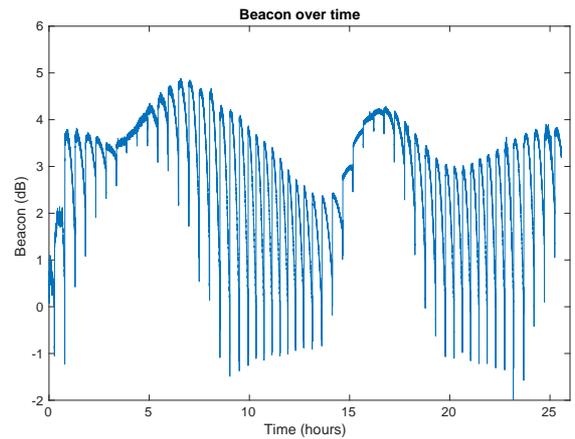}
  \caption{Beacon signal level over time. \label{fig:beacon}}
\end{figure}

Figure \ref{fig:beacon} shows the beacon signal level over time. The lowest
beacon level in our setup is around $-24$~dB. In this test we achieved a mean
beacon level of $2.89$~dB, and a standard deviation of $1.15$~dB. The beacon
level reduction happens because the antenna remains stopped until the next
tracking cycle. The tracking algorithm then restores the beacon level to a
higher value. Variations in the upper level are due to weather conditions,
temperature and time of day.

\section{Conclusions}

The results obtained show that the least squares technique implemented in the
real antenna control system is able to track the Brasilsat B3 satellite
satisfactorily, confirming the work of \cite{Laine2000}.
Also, the adaptation described in this paper with the use of RLS performed well,
removing the need for matrix inverses and increasing the adaptability of the
model due to the forgetting factor.

\section*{Acknowledgements}

The authors would like to thank BrasilSAT Harald S/A fortheir support during the
project development.

\bibliographystyle{plain}

\bibliography{sat-track}

%\begin{thebibliography}{99}
%
%\bibitem{ref1} L. Lamport, \textit{A Document Preparation System: \LaTeX, User's Guide and Reference Manual}. Addison Wesley Publishing Company,
%1986.
%\bibitem{ref2} F. C. Silva and J. J. Sousa, ``This reference is just an example," ~\textit{Journal of Examples}, v. 5, pp. 52--55, May 1999.
%\end{thebibliography}

\end{document}